\documentclass[a4paper]{aa}
\usepackage{graphicx}
\usepackage{txfonts}
\usepackage{natbib}
\bibpunct{(}{)}{;}{a}{}{,} 
\usepackage{bm}

\begin{document}
\title{Characterizing the velocity field in hydrodynamical simulations of 
	   low-mass star formation using spectral line profiles}
\titlerunning{Characterizing the velocity field in simulations of 
			  low-mass star formation using spectral lines}

\author{
  C.~Brinch\inst{1} \and 
  M.~R.~Hogerheijde\inst{1} \and 
  S.~Richling\inst{2}}
\date{}
\institute{
  Leiden Observatory, Leiden University, 
  P.O.~Box 9513, 2300 RA Leiden, The Netherlands\\ 
  \email{brinch@strw.leidenuniv.nl} \and 
  Institut d'Astrophysique de Paris, 
  98 bis Boulevard Arago, 75014 Paris, France}

\abstract
{When low-mass stars form, the collapsing cloud of gas and dust goes through 
 several stages which are usually characterized by the shape of their spectral 
 energy distributions. Such classification is based on the cloud morphology 
 only and does not address the dynamical state of the object.}
{In this paper we investigate the initial cloud collapse and subsequent disk
 formation through the dynamical behavior as reflected in the sub-millimeter 
 spectral emission line profiles. If a young stellar object is to be 
 characterized by its dynamical structure it is important to know how 
 accurately information about the velocity field can be extracted and which 
 observables provide the best description of the kinematics. Of particular 
 interest is the transition from infalling envelope to rotating disk, because 
 this provides the initial conditions for the protoplanetary disk, such as mass 
 and size.}
{We use a hydrodynamical model, describing the collapse of a core and formation 
 of a disk, to produce synthetic observables which we compare to calculated 
 line profiles of a simple parameterized model. Because we know the velocity 
 field from the hydrodynamical simulation we can determine in a quantitative 
 way how well our best-fit parameterized velocity field reproduces the 
 original. We use a molecular line excitation and radiation transfer code to 
 produce spectra of both our hydrodynamical simulation as well as our 
 parameterized model.}
{We find that information about the velocity field can reasonably well be 
 derived by fitting a simple model to either single-dish (15$''$ resolution) 
 lines or interferometric data (1$''$ resolution), but preferentially by using 
 a combination of the two. The method does not rely on a specific set of 
 tracers, but we show that some tracers work better than others. Our result 
 shows that it is possible to establish relative ages of a sample of young 
 stellar objects using this method, independently of the details of the 
 hydrodynamical model.}
{}
\keywords{Radiative transfer -- Hydrodynamics -- Stars: formation -- 
          ISM: molecules -- Line: profiles}
\maketitle

\section{Introduction}\label{intro}
Low-mass young stellar objects (YSOs) are well-studied phenomena. 
Observationally, these objects have been measured in many different wavelength 
regimes in order to determine intrinsic properties such as their mass, 
luminosity, mass accretion rate, etc. YSOs are traditionally classified by the
shape of their spectral energy distribution (SED)~\citep{lada1984}, which 
reflects their general morphology. Subsequently, the envelope and disk 
densities, temperatures and chemistry can be studied using millimeter 
continuum and molecular line measurements. All of these derived quantities have 
been used to piece together a consistent chronology of the process of star 
formation and to establish a reference from which the age of a YSO can be 
reliably obtained~\citet{mardones1997,gregersen2000}. 

A theoretical evolution scenario of YSOs has been established for many years
\citep[e.g.,][]{ulrich1976,shu1977,cassen1981,terebey1984,bodenheimer1990,
basu1998,yorke1999}. In these models (with the exception of the Shu model, 
which is spherically symmetric) a circumstellar disk is formed, after the 
collapse of the initial cloud sets in, due to conservation of angular 
momentum. It is therefore natural to use the kinematic properties of the models 
to characterize the evolutionary stage of the object, because the distribution 
of angular momentum changes in a monotonous way as the cloud contracts and 
matter spins up. Unfortunately, measuring the velocity field observationally is 
not such a simple matter. Kinematic information can only be derived from 
spectral emission lines and there is no way to directly solve for the velocity 
field from measurements of such lines. The reason for this is that the spectral 
profile depends on a range of physical parameters, including the local density, 
temperature, molecular abundance, and turbulence in the region where the 
transition in consideration is excited. All of this can be taken into 
consideration and self-consistent models, including a general parameterized 
velocity field, can be built and fitted to the observed spectra. The question 
is, however, how reliable the derived velocity field is given the input model, 
and to what extent the derived velocity fields are consistent with the 
prediction of the models.

The velocity field of YSOs has been well studied observationally in the last 20 
years. At first, mainly infall were observed \citep[e.g.,][]{calvet1992,
vanlangevelde1994}, but soon afterward, objects showing a mix of infall and 
rotation was discovered \citep{Saito1996,Ohashi1997,Hogerheijde2001,
belloche2002}. More recently, protoplanetary disks have been studied in very 
high resolution using sub-millimeter interferometry and many of these have 
velocity fields with rotation only \citep[e.g.,][]{Lommen2008}. The common 
interpretation of these observations is that the objects showing infall only 
are early-type embedded objects, the rotating disks are the end product of the 
collapse, and the objects showing both infall and rotation are in some kind of 
transition. Initial modeling work using a parameterization for the velocity 
field has been done by \citet{myers1996} for low-mass star formation and by 
\citet{keto1990} for the formation of high-mass stars. However, it has never 
been attempted to quantify the degree of the transition, how far the object has 
passed from the embedded phase to the protoplanetary disk phase.

In this paper we will address this question, by comparing synthetic spectra,
calculated from a hydrodynamical simulation of a collapsing cloud, to a model
with a simple parameterized velocity field. The comparison is done in spectral 
space in the same way as one would compare a model to real observations. By
considering different transitions of different molecules in two different
resolution settings, we investigate how to obtain the most reliable 
characterization of the velocity field. The importance of being able to 
reliably model the observations of YSOs using a simple generic model becomes 
evident when considering the capabilities of the upcoming ALMA (Atacama Large 
Millimeter Array). While it is possible to build sophisticated custom models, 
or even directly use the hydrodynamical simulation as a template model to 
describe single objects, this is a very time-consuming process. ALMA will 
produce very high resolution observations ($\sim$ 0.01$''$) in snapshot mode 
and in order to deal with such vast samples of YSOs, simple yet reliable models 
are needed.

The layout of this paper is as follows: In Sect.~\ref{model} we present the 
hydrodynamical simulation of~\citet{yorke1999} that we use to make synthetic 
observations and describes our model and the way we fit this to the synthetic 
spectra. Section~\ref{lines} presents the synthetic observables and also a few 
analytical examples of idealized cases for comparison. Section~\ref{results} 
shows our results of the spectral comparison and the derived velocity fields, 
followed by a discussion and conclusion in Sects. \ref{discussion} and 
\ref{conclusion}, respectively.

\section{Simulations and models}\label{model}
\subsection{Hydrodynamical collapse}
For the description of the gravitational collapse, we use the hydrodynamical
scheme of~\citet{yorke1999}. In particular we use the model which simulates the
formation of an J-type star given an initial cloud mass of 1 M$_\odot$ and 
radius of 6667 AU (1$\times10^{15}$ m). The cloud is initially isothermal at 10 
K and is given a constant solid body rotation of 1 $\times$ 10$^{-13}$ 
s$^{-1}$. In this model, an equilibrium disk is formed after about 75000 years 
after the onset of the collapse, which then grows up to several thousands of AU 
as angular momentum is transported outwards. The disk which is formed resembles 
a thin, flaring disk with a pressure scale height at 500 AU from the star of 
$h = 0.15R$ and $h = 0.19R$ at 1000 AU from the star.

The code used to calculate the collapse is described in detail by
\citet{rozyczka1985},~\citet{yorke1993},~\citet{yorke1995}, and 
\citet{yorke1999}, but a brief overview is provided here as well. The standard 
equations of hydrodynamics including radiative transfer and the Poisson 
equation for the gravitational potential is solved on five regular nested grids 
of 68 by 68 cells. The grid does not resolve the central star, but the 
centermost grid cell is treated as a sink into which mass and angular momentum 
can flow. The stellar luminosity is calculated from the mass and mass accretion 
of this cell and the luminosity is used as a boundary condition for the 
radiative transfer calculations. The radiative transfer is solved using the 
frequency dependent flux-limited diffusion method~\citep[see][for the details]
{yorke1999}, in order to obtain a self-consistent temperature distribution. 
Artificial viscosity is included in order to deal with shocked regions and 
physical viscosity is provided by an $\alpha$-prescription~\citep{shakura1973} 
to include angular momentum transport in the disk. 

\begin{figure}
  \begin{center}
	\includegraphics[width=8.5cm]{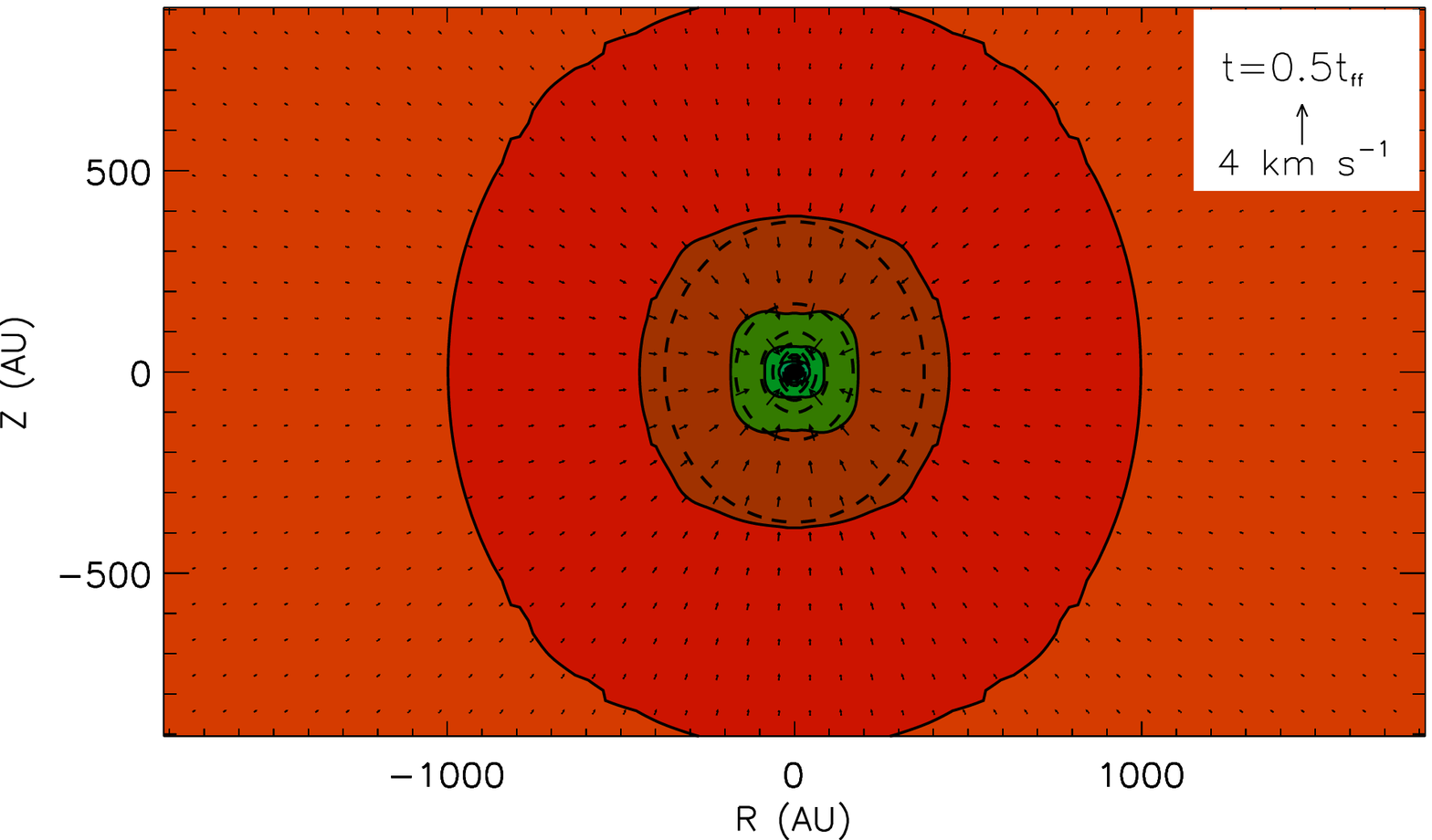}\\
	\includegraphics[width=8.5cm]{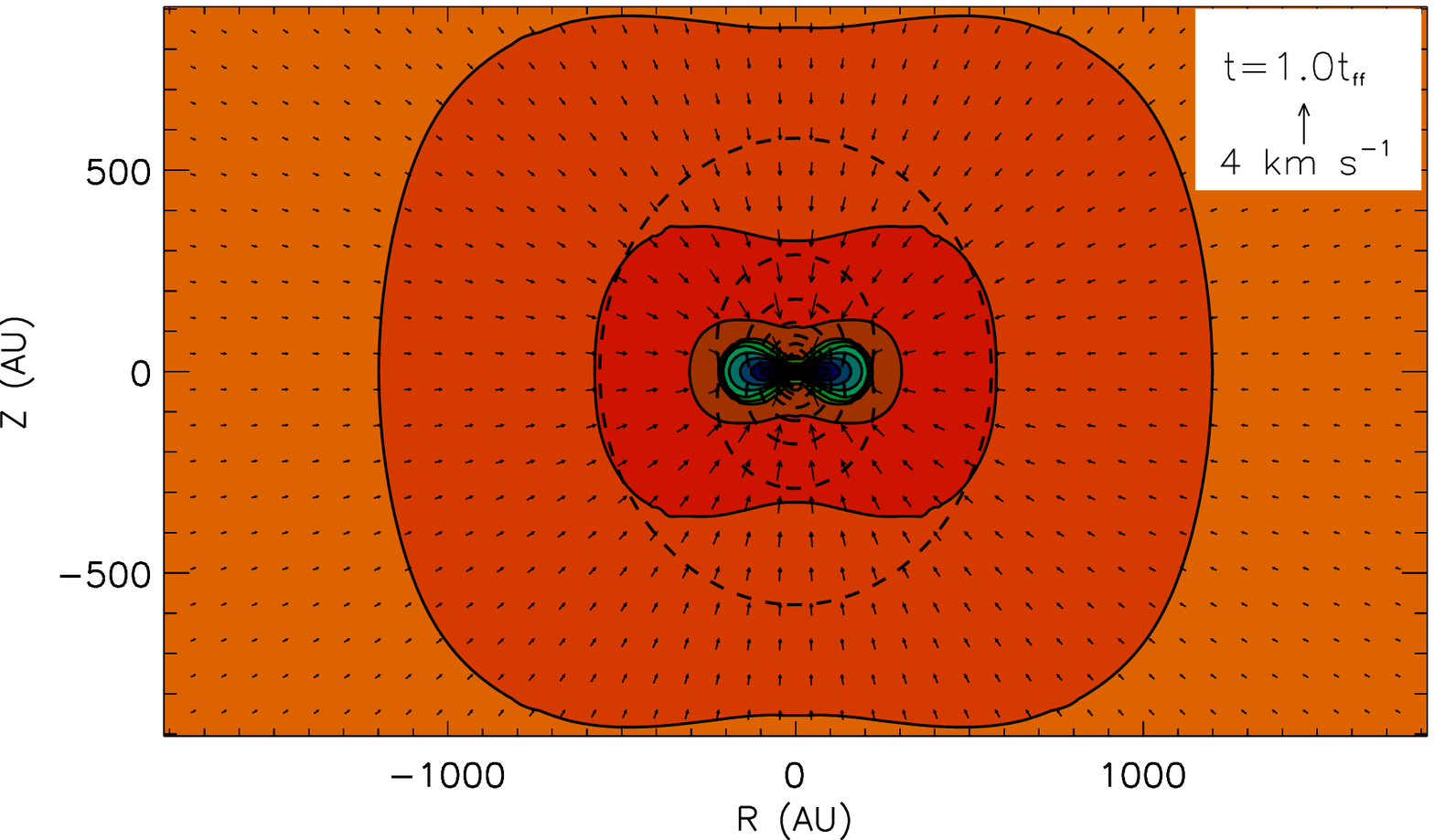}\\
	\includegraphics[width=8.5cm]{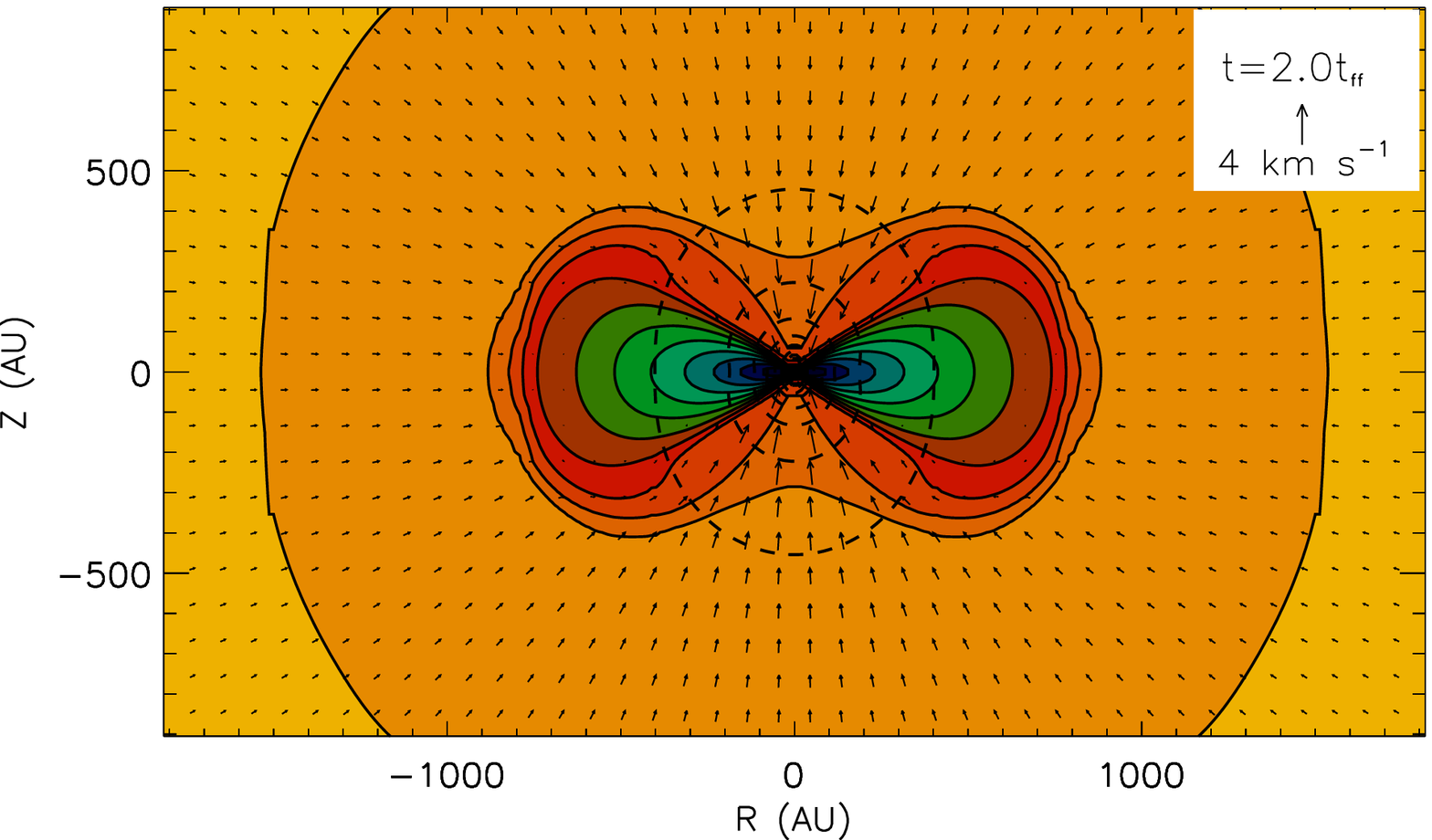}
  \end{center}
  \caption{Model structure at three different ages. Full contours are the
  density starting at 3$\times$10$^5$ cm$^{-3}$ and increasing with $\log \rho$
  = 0.5. The temperature is shown as dashed contours, starting at 20 K and
  increasing with 10 K. The arrows show the infall component of the velocity
  field.}\label{models}
\end{figure}

We use the initial free fall time scale to describe the age of the simulated 
YSO. Free fall time is given as
\begin{eqnarray}
t_{ff}=\sqrt{\frac{\pi^2R^3}{8GM}},
\end{eqnarray}
which for this particular model is approximately 1$\times 10^5$ years. The 
simulation can roughly be divided into three parts, namely the infall dominated 
pre-disk stage at 0.5 t$_{ff}$, the disk formation stage at about 1.0 t$_{ff}$ 
in which both infall and rotation are important, and the rotation dominated 
disk growth stage at 2.0 t$_{ff}$. The entire simulation runs for about 2.5 
t$_{ff}$ at which point it halts. Representative snapshots of the density, 
temperature, and velocity structure at these three stages are shown in 
Fig.~\ref{models}. In the three panels, the temperature and density contours 
are kept at the same values for comparison. 

We have selected a number of snapshots throughout the simulation, 13 in total
evenly spaced by about 0.2 in free-fall time, as input for our molecular 
excitation and frequency dependent radiation transfer code \emph{RATRAN}
\citep{hogerheijde2000a}. This code runs on a similar, but coarser grid, so the
density, temperature and velocities of the hydrodynamical simulation can 
directly be remapped onto a \emph{RATRAN} grid. The level population in each 
cell is then solved iteratively by propagating photons randomly through the 
grid until the level populations are in equilibrium with the radiation field. 

The reason why we do not use every snapshot (several hundreds, depending on the 
time resolution in the output) is that the molecular excitation calculations 
are quite time consuming. The CPU time required to determine the level 
populations for a single snapshot is comparable to the CPU time it takes to run 
the entire hydrodynamical simulation, which is the reason why such accurate 
radiation transfer methods are not included in the simulation itself. However, 
the level populations and hence the line profiles do not change dramatically 
from one time step to the next, so there is actually no need for our purpose to 
use more than one snapshot per 0.1 -- 0.2 t$_{ff}$.

In \emph{RATRAN}, a molecular abundance is added to the model reflecting the 
chemical abundance of the molecule in question. The simplest possible abundance 
description uses a constant average abundance relative to the H$_2$ density 
which is how we populate the models with molecules in this paper. More 
realistic abundance profiles, based on temperature and density dependent 
freeze-out and desorption rates, have been discussed by, e.g.,~\citet{lee2004} 
and~\citet{jorgensen2004}. In Sect.~\ref{discussion} we discuss the effects on 
our results, if a more realistic abundance distribution is added to the 
synthetic observations. For an in-depth treatment of the inclusion of depletion 
chemistry in the hydrodynamical simulation and its effect on the resulting 
spectra, we refer the reader to the subsequent paper by Brinch et 
al.~\emph{(in prep.)}.

We execute separate \emph{RATRAN} calculations for different molecular species. 
As mentioned above, these calculations are relatively time consuming, so we 
have chosen to focus on only three commonly observed molecules, CO, HCO$^+$, 
and CS, and their isotopologues. These three species probe different densities 
and have different optical depth and therefore the spectral profiles come out 
very different even for the same underlying snapshot. We can of course repeat 
the entire exercise for any molecule for which collisional rates are available 
and this makes our method very flexible. The collisional rates used in this 
paper come from the \emph{LAMDA} database\footnote{available at 
http://www.strw.leidenuniv.nl/$\sim$moldata} \citep{schoier2005}. When the 
level populations have converged, the model is ray-traced in frequency bands 
around the desired transitions to obtain spectral image cubes from which 
spectral lines can be extracted. In the following we consider increasingly 
higher transitions for the three species, so that these not only probe 
increasingly higher densities but also increasingly higher temperatures.

We do not claim that this hydrodynamical collapse model provides a particularly 
realistic description of nature. Neither do we try to fit this model to 
observation, although in principle, the method presented in this paper can be 
used to compare and constrain hydrodynamical models with molecular line 
observations. Instead, we employ the simulation to provide a time resolved 
velocity field topology going from an infall dominated field to a rotation 
dominated field. The scope of this paper is to show how the velocity field is 
reflected in the emission line profiles and to see to what extent the velocity 
field can be reconstructed from these lines. As long as the hydrodynamical 
simulation approximates reality, our method is valid. The two main caveats of 
our simulation are the lack of any kind of mass loss through either a wind or 
and outflow and the unknown initial condition for the angular momentum 
distribution. The lack of mass loss means that all the mass that is initially 
in the model ends up in either the disk or the star, and especially the disk 
ends up being very large and massive. An outflow would not only carry away mass 
but also angular momentum and the result of this would be a smaller, less 
massive disk. The effect of the second item, the initial angular momentum 
distribution is somewhat more uncertain and a point of further study, which 
however is beyond the scope of this paper.

\subsection{Parameterized model}
We have chosen a particularly simple parameterization for the model which we 
use to derive the velocity field. This model is a simplified version of the one 
used by~\citet{brinch2007} to interpret single dish observations of the YSO 
L1489~IRS.  

In this paper we use a spherical version of this model, where both the density 
and the temperature is given by power-laws
\begin{eqnarray}
  n &=& n_0(r/r_0)^{-p}, \\
  T &=& T_0(r/r_0)^{-q},
\end{eqnarray}
which means that our parameterized model neglects the flattening of the disk. 
We assume that the inclination is known to at least 10-20\%. Although our model 
is spherically symmetric, it still has a rotation axis, which may be inclined 
with respect to our line of sight. For the remainder of this paper we adopt an
edge-on view of the simulation ($i$=90$^\circ$). The velocity field is 
described by two free parameters, the central mass $M_*$ and an angle $\alpha$, 
that is defined as,
\begin{eqnarray}\label{alpha}
  \alpha = \arctan \left(\frac{v_r}{v_\phi}\right),
\end{eqnarray}
where $v_r$ and $v_\phi$ are the free-fall and Keplerian velocity components, 
respectively, given by
\begin{eqnarray}
  v_\phi cos(\alpha) = -\frac{v_r}{\sqrt{2}} sin(\alpha)= \sqrt{\frac{GM_*}{r}}.  
\end{eqnarray}
The angle $\alpha$ is thus the angle of the resulting velocity field vectors 
with respect to the azimuthal direction, and therefore the ratio of infall to 
rotation. In other words, if $\alpha=0$ the cloud is in pure Keplerian motion 
whereas if $\alpha=\pi/2$ the cloud is collapsing in free fall. This 
parameterization is allowed because both rotation and infall have a 
$\sqrt{M_*/r}$ dependency.

The advantage of this simple parameterization is that it has few free 
parameters and, more importantly, for a large part of the hydrodynamical 
simulation the density and temperature are actually very well described by 
power-laws. The disadvantage is of course that it cannot describe the disk 
structure at all. However, since we are aiming at characterizing the velocity 
field, the exact shape and distribution of disk material is less important and 
it turns out that is has little impact on our result as long as we fit our 
density model to the mid-plane density of the hydrodynamical output.

The $\alpha$ parameter has no direct equivalent in the hydrodynamical 
simulation because the ratio of infall to rotation is not generally constant 
with radius. In the remainder of this paper we do however use an averaged value 
for $\alpha$ where we have radially averaged the individual $\alpha$ values 
along the mid-plane in order to evaluate the quality of our results.

In this paper we assume that the inclination is known (we consider only the 
edge-on case, but our result holds true except for inclinations very close to
face-on). In the beginning of the simulation, where rotation is less important,
the inclination is harder to constrain observationally, but on the other hand, 
it does not really matter since the object is mostly spherical anyway. As soon 
as a disk has formed, the inclination can easily be constrained by modeling the 
SED or from near-infrared imaging. In any case, if the inclination is not 
known, it can in principle be included as a free parameter at the cost of 
computation time. 

A final caveat of this simple model is the flat abundance distribution. As 
discussed above, the abundance distribution will evolve dynamically with the 
evolution of the temperature and density, and this will have an effect on the 
spectral lines. This effect can be taken into account in the modeling using the
approach of~\citet{jorgensen2004}, but for the sake of clarity we will 
disregard the chemistry in the main part of this paper. 

\subsection{Obtaining the best fit}\label{pikaia}
When fitting the parameterized model to the hydrodynamical solution, we do so
in a sequence of comparisons between different synthetic observables. First
we constrain the five parameters $n_0, T_0, p, q,$ and $R_{out}$ ($R_{out}$ is 
the outer radius of the model beyond which the power-laws are truncated) in the 
density and temperature model by fitting radial continuum profiles at 450 and 
850 $\mu$m. A synthetic continuum image is calculated by \emph{RATRAN} at each 
wavelength based on the hydrodynamical solution. Similarly we calculate images 
for the parameterized model that is then evaluated with respect to the image 
from the hydrodynamical solution by calculating the $\chi^2$ measure. In other 
words, we obtain the set of parameters which gives the best representation of 
the hydrodynamical structure in the image space. The wavelengths and image 
resolution are chosen so that the images mimic real observations by 
instruments such as SCUBA on the James Clerk Maxwell Telescope (15 meter single 
dish, with a typical resolution of 15$''$, corresponding to $\sim$2000 AU at 
the distance of the Taurus star forming region).

When the physical structure has been determined we fix the molecular abundance
by fitting an emission line of an optically thin isotope. In the case of 
HCO$^+$, this isotope would for example be H$^{13}$CO$^+$, whereas for CO we 
could use C$^{17}$O. An optically thin line traces the total column density and 
can as such be used to derive the average abundance. Since we have not yet 
determined the velocity field, we aim at fitting the integrated intensity 
rather than the peak intensity or the line profile itself. The procedure is the 
same as before: A single spectrum towards the center is calculated by 
\emph{RATRAN} based on the hydrodynamical solution which is then used as a 
synthetic observation to fit the same spectrum based on the parameterized 
model. Again, we use a spectral and spatial resolution which is typical for 
existing single dish telescopes (15$''$).

Finally, when the abundance is fixed, we fit the velocity field parameters.
This is done by comparing optically thick lines. In the case where we 
simulate single dish observations, we use only a single spectrum towards the 
center of the object, in which case we fit only the central mass and a single
angle $\alpha$. If we use simulated interferometric observations, where there
are typically several resolution elements across the object, we fit spectra 
which are separated by one synthesized beam, with as many corresponding angles
$\alpha$.

In order to fully explore the parameter space to reach the optimal solution,
we use the genetic algorithm \emph{PIKAIA}~\citep{charbonneau1995} which is a
stochastic optimizer. This algorithm is applied in each step describe above 
and is very successful in finding the true minimum of the search space. The
benefit of using this algorithm for this purpose is that it takes a relative
low number of model calculation to reach the optimal solution compared to the 
more traditional ``brute force'' method where a grid of models are calculated.
Since calculating a model spectrum takes of the order minutes to complete the
applicability of our method is highly dependent of the number of models which
are required to be calculated. Another benefit of using a code such as 
\emph{PIKAIA} is that the entire parameter space gets well sampled which 
enables us to determine the $\chi^2$ distribution along each parameter axis. 
The resulting distributions are approximated by Gaussians, and the FWHM of 
these are used to get a handle on the robustness of the fit. It should be noted 
that this is not an error bar in the sense that our model fit is consistent 
with the synthetic data if it falls within the range of the FWHM of the 
Gaussian approximation to the $\chi^2$ distribution. These numbers only show 
how sharply defined the minimum of the parameter space is for a specific 
optimization. 

\begin{figure}
  \begin{center}
	\includegraphics[width=8.5cm]{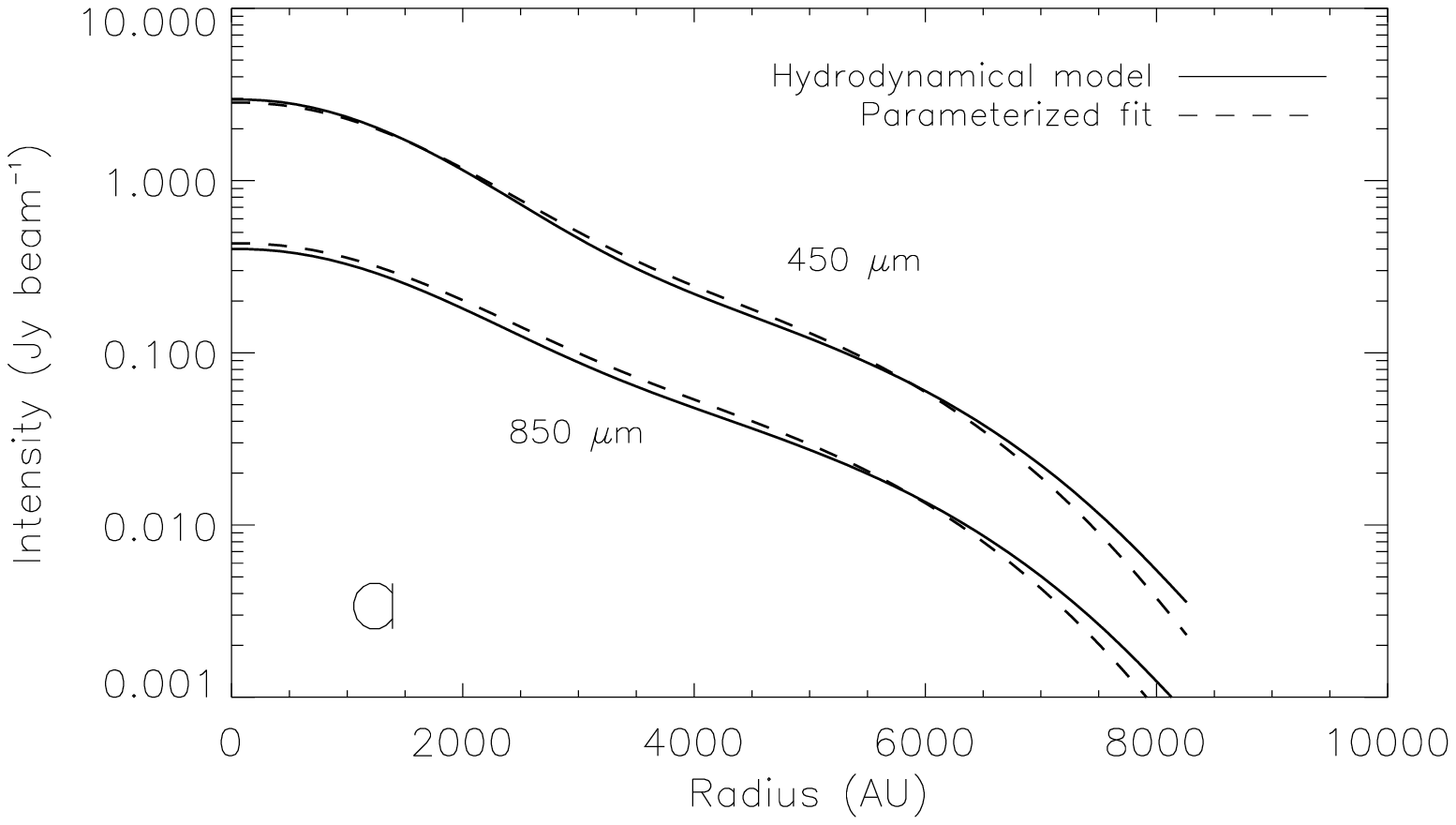}\\\
	\includegraphics[width=8.5cm]{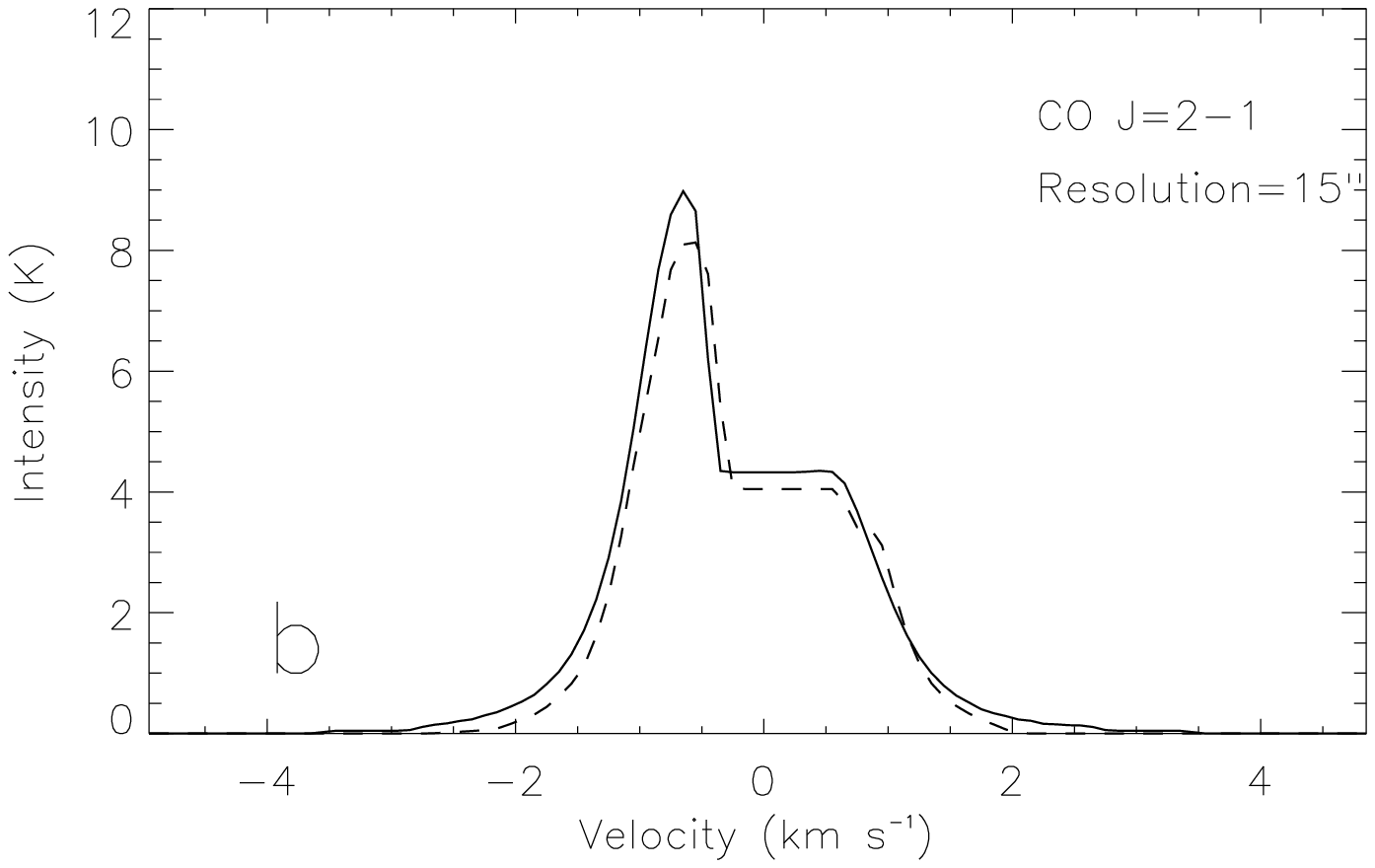}\\
	\includegraphics[width=8.5cm]{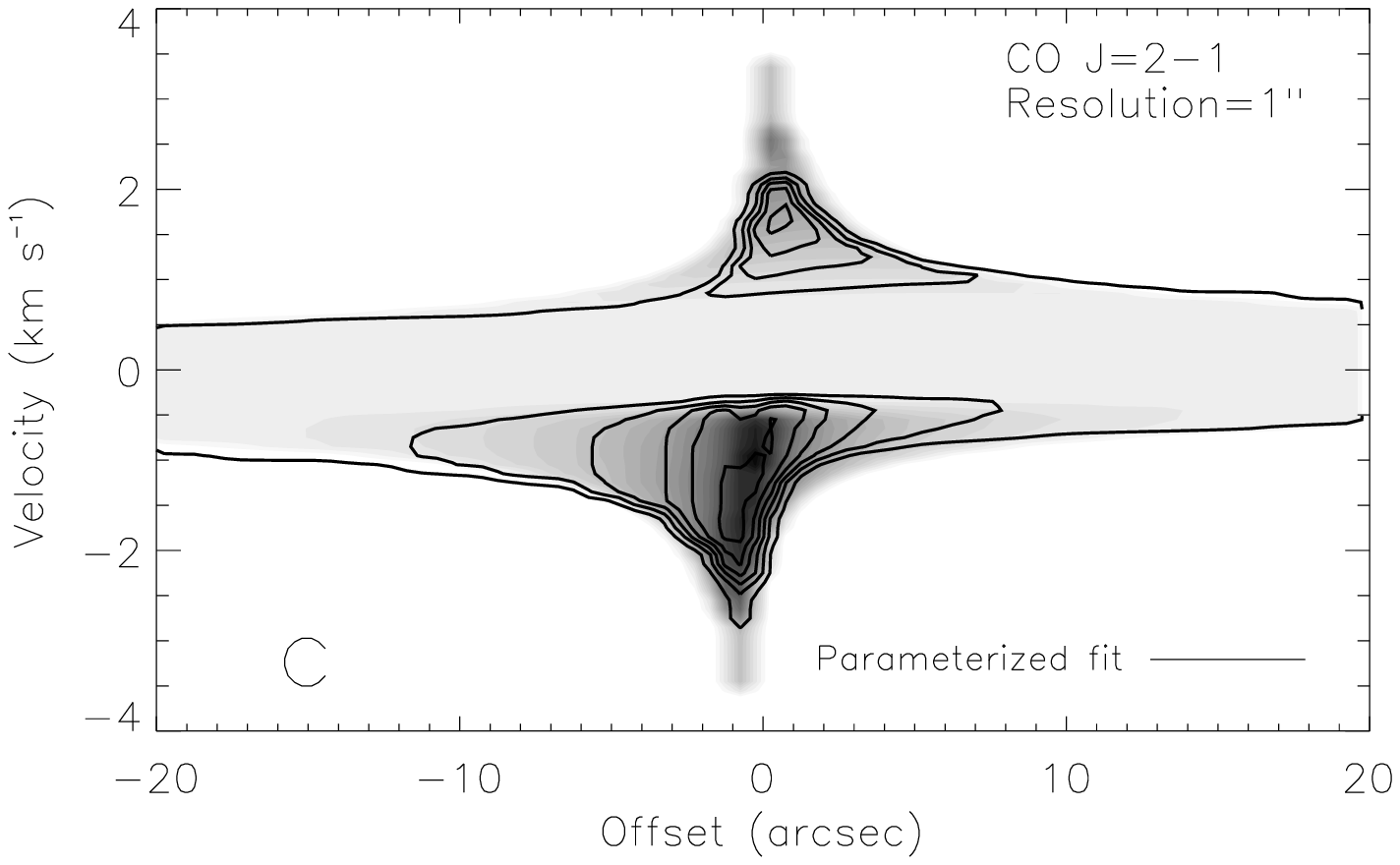}
  \end{center}
  \caption{Example of a fit by our parameterized model to the hydrodynamical
           solution at t=1.0 t$_{ff}$ for the case of CO. The full lines are
		   the output of the hydrodynamical simulation and the dashed lines are
		   the fit by our parameterized model. In panel C, the hydrodynamical
		   output is shown by the gray scale and the model fit by contour lines.
		   }\label{testfits}
\end{figure}

Figure~\ref{testfits} shows how well we are able to approximate the 
hydrodynamical solution by our simple parameterized model. The example shown
is just a randomly chosen snapshot (t=1.0 t$_{ff}$). 

\section{Line profiles and the velocity field}\label{lines}
Initially the cloud collapse model is given a solid body rotation. As 
gravitation sets in and infall velocity builds up towards free fall, matter is
redistributed and so is the angular momentum. Material ends up in the plane of
rotation, where it spins up, approaching the Keplerian velocity. At this point,
the matter cannot be given a higher angular momentum, so as matter continuously
moves inward from outside of the disk, angular momentum is transported outwards 
through the disk resulting in growth of the disk. 

\begin{figure}
  \begin{center}
	\includegraphics[width=8.5cm]{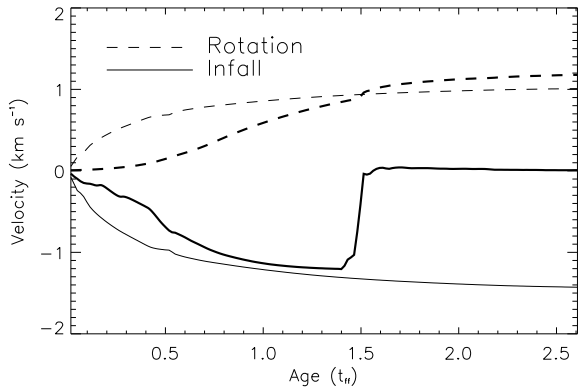}
  \end{center}
  \caption{The rotation and infall velocities of the collapse simulation in the 
  		   mid-plane at a radius of 500 AU plotted as a function of free-fall 
		   time. The velocities of the simulation are shown in thick lines.
  		   In this figure we also show the Keplerian and free-fall velocities 
		   which are both functions of the central mass. These are shown as 
		   thin lines.}\label{velocity}
\end{figure}

\begin{figure}
  \begin{center}
	\includegraphics[width=8.5cm]{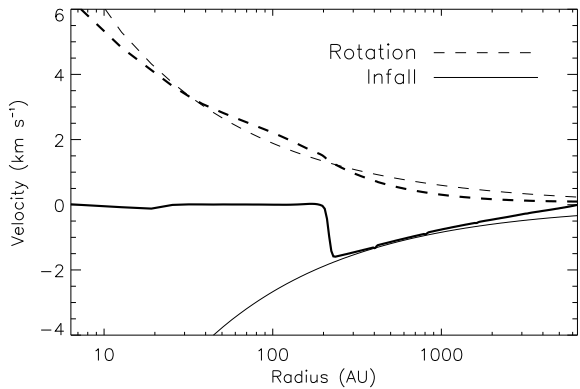}
  \end{center}
  \caption{The rotation and infall velocities in the simulation along the
  	       mid-plane at an age of t=1.0 t$_{ff}$. As in Fig.~\ref{velocity},
		   the thin lines correspond to free-fall and Keplerian velocity
		   profiles.}\label{velocity2}
\end{figure}

In Fig.~\ref{velocity} the radial and the azimuthal components of the velocity
field in the disk mid-plane at a radius of 500 AU are plotted. Also shown in 
this plot is the free-fall and Keplerian velocities calculated as a function of 
the central core mass which grows in time as mass is being accreted onto the 
star. Initially, the radial velocity builds up faster than the azimuthal 
velocity component, but at an age of t=1.5 t$_{ff}$ the disk expansion shock 
wave passes 500 AU and afterward the region is completely dominated by 
rotational velocity close to the Keplerian. Indeed the velocity is seen to be 
slightly super-Keplerian after the expansion wave passes because of the 
outwards angular momentum transport. Similarly, Fig.~\ref{velocity2} shows 
again the infall and rotation velocity profiles, but for a single time slice 
(t=1.0 t$_{ff}$). In this figure, the velocities are plotted against the 
mid-plane radius and also shown are the free-fall and Keplarian profiles, 
indicated by thin lines. The disk radius is clearly seen by the break in the 
infall profile around 200 AU. It is also seen that the velocity field is very 
well approximated by free-fall outside of the disk radius, whereas the inside 
of the disk radius is well described by Keplerian motion.

The velocity of a gas, is reflected in the shape of the emission line profiles. 
This includes both the random particle motion and the systematic motion due to 
potential gradients. For the model cloud considered here, the former is of 
minor importance although we do include a microturbulent field with a Gaussian
FWHM of 0.2 kms$^{-1}$ in our calculations. The systematic motion is of course 
our main interest here, because this is what governs the dynamics of the cloud.

Initially, the cloud is collapsing in a nearly spherical symmetric way which is
entirely dominated by radial motions. This should be reflected in the line 
profiles as the characteristic infall asymmetry. In the case of optically thick
lines, the difference in temperature of the red shifted and the blue shifted 
peaks provides a measure of the optical depths~\citep[e.g.,][]{evans1999}. In 
the latest stage, when most of the gas has been accreted onto the disk which is 
dominated by rotation, the line profiles are expected to be double peaked, with 
both peak having the same intensity. At this point, gas has reached such a low 
density outside of the disk, that we do not expect to see any contribution to 
the line from radial motions. In the intermediate stage however, where the disk 
is still accreting significantly, both velocity components, radial and 
azimuthal, will contribute to the spectral line profile. The exact shape of 
such a profile is hard to predict because it is critically dependent on the 
exact optical depth, the critical density of the transition, the spatial 
resolution, and the system inclination. This is further complicated by the fact 
that no analytical model exists of the transition from a collapsing envelope to 
a viscously supported protoplanetary disk. 

\begin{figure}
  \begin{center}
	\includegraphics[width=8.5cm]{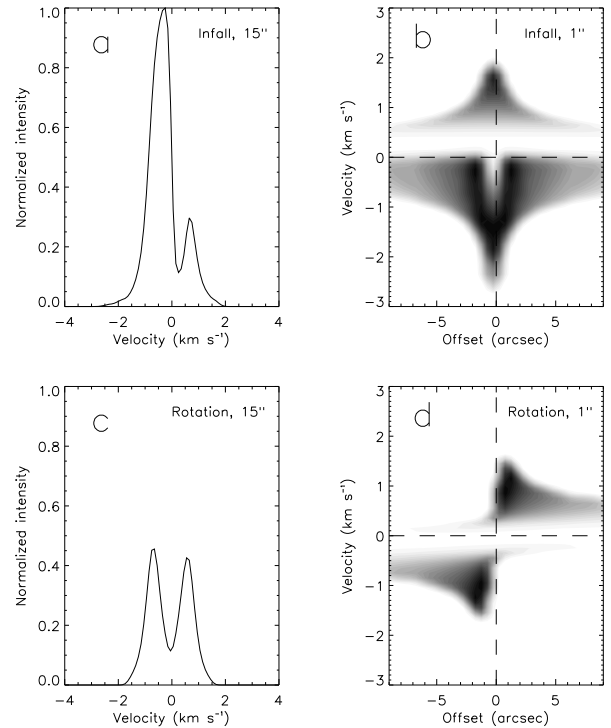}
  \end{center}
  \caption{Low-resolution spectra and high-resolution position-velocity 
  		   diagrams of a purely (free) infalling model and a purely 
		   (Keplerian) rotating model.}\label{ideal}
\end{figure}

As an illustration, we show model spectra in Fig.~\ref{ideal} which have been 
calculated with a model which is either in free fall (with no azimuthal 
velocity component at all) or in Keplerian rotation (no radial components). The 
top-most panels in this figure show the infalling models in which the 
infall-asymmetry~\citep{evans1999} is clearly seen. The corresponding 
position-velocity diagram also resembles the CO 2--1 at t=0.5t$_{ff}$ panel in 
Fig.~\ref{pvs}. Similarly, the PV-diagram in the lower row of Fig.~\ref{ideal} 
resembles the late stage PV-diagram in Fig.~\ref{pvs} and the low-resolution 
spectrum shows the double-peak feature which is associated with rotation.
Therefore, if the gas has the particular density in which an observed 
transition is strongly dominated by one or the other type of motion, it is 
possible to tell the difference from the shape of the line or the intensity 
distribution of the PV-diagram. This, however is rarely the case because the 
velocity field in general changes with increasing density (due to angular 
momentum conservation) and in most cases, the gas has non-vanishing velocity 
components in all three spatial directions.

\subsection{Low resolution spectra}\label{subsec:lowres}
\begin{figure}
  \begin{center}
	\includegraphics[width=8.5cm]{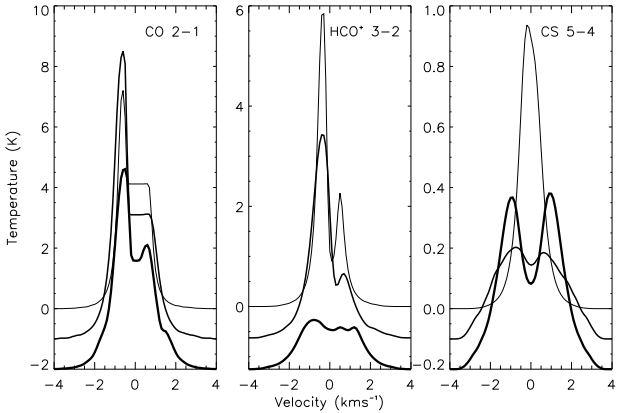}
  \end{center}
  \caption{Spectral line profiles of three different tracers at three different
  		   ages. t$_{ff}$ = 1.0 and 2.0 are offset by $-10\%$ and $-20\%$ 
		   respectively. These spectra have been convolved with a 15$''$ 
		   Gaussian beam to resemble typical single-dish telescope 
		   observations.}\label{specs}
\end{figure}

Using our hydrodynamical simulation as input model we can calculate spectra of
any disk and envelope configuration ranging from purely collapsing to entirely 
rotating. In Fig.~\ref{specs} is shown the result of such a calculation. These
spectra are convolved with a 15$''$ FWHM Gaussian beam so that they are 
comparable to observed spectra obtained with a typical large single dish 
telescope. The system inclination has been chosen to be exactly edge-on so that 
the viewing angle is the same as in Fig.~\ref{models}.

In the leftmost panel of Fig.~\ref{specs} are shown CO $J=$ 2--1 spectra based
on the three time snapshots from Fig.~\ref{models}. The CO lines are seen to be
very optically thick throughout the simulation which is a result of the 
relatively high CO abundance. Actually, the optical depth of the CO lines is so 
high that we do not even see the rotating disk in the last snapshot, where the 
profile still bears the typical infall signature also seen in the two earlier 
snapshots. Even though the densities are very low ($< 10^4$ cm$^{-3}$) outside 
of the disk at t=2.0 t$_{ff}$, there is still enough column to entirely mask 
the disk. In the middle panel, HCO$^+$ $J=$ 3--2 spectra are shown. These are 
optically thick too, but not as much as the CO lines. The infall asymmetry is 
very pronounced in the t=0.5 t$_{ff}$ spectrum and is still visible in the 
t=1.0 t$_{ff}$. In the last snapshot, the infall signature is gone, and we only 
see a flattened Gaussian-like spectrum. In the right-most panel of 
Fig.~\ref{specs} we have plotted CS $J=$ 5--4. This line traces very high 
densities ($> 10^6$ cm$^{-3}$). Notice that the intensity of these lines are 
about an order of magnitude lower that the intensity of the CO lines. In the 
earliest snapshot, this transition is actually optically thin and only a hint 
of asymmetry is seen here. As we go to later stages, the CS lines also start to 
become optically thick. This line, however, clearly displays a double peaked 
profile in the late stage. We also see, contrary to the CO and HCO$^+$ lines 
that the width of the CS spectra increases with time.

\begin{figure}
  \begin{center}
	\includegraphics[width=8.5cm]{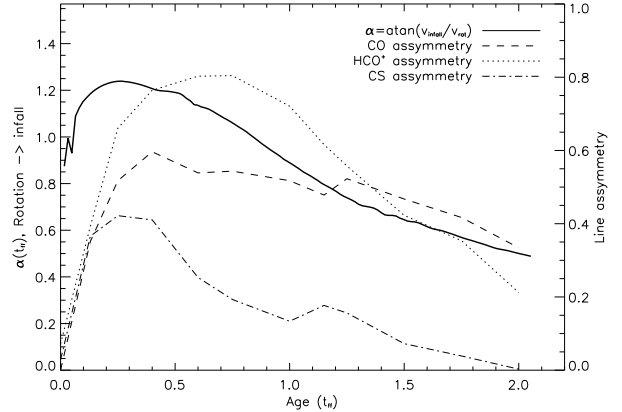}
  \end{center}
  \caption{The averaged $\alpha$ from the hydrodynamical simulation plotted against 
  free-fall time (thick solid line). The punctured lines show the dimensionless
  asymmetry of the single-dish spectral lines as a function of time.}
  \label{assym}
\end{figure}

To illustrate how the shape of the line profile is a reflection of the 
underlying velocity field, we have quantified the line asymmetry by measuring
the intensity difference between the two peaks, normalized with the peak 
intensity of the line. This gives us a dimensionless number between 0 and 1,
where 0 corresponds to the extreme asymmetric line and 1 is a perfectly 
symmetric line. We have plotted this quantity for the whole time series of 
model spectra of CO, HCO$^+$, and CS in Fig.~\ref{assym}. These line 
asymmetries are plotted as dotted lines. The full line in Fig.~\ref{assym}
is the radially averaged angle $\alpha$ from the hydrodynamical simulation as
calculated by Eq.~\ref{alpha}. The Pearson correlation coefficients between 
$\alpha$ and line profile asymmetry are 22\%, 61\%, and 83\% for CO, HCO$^+$, 
and CS, respectively, meaning that the asymmetry of the CS lines is a quite 
well correlated with the ratio of infall to rotation. Similarly, the central 
stellar mass (as given by the hydrodynamical simulation) can be correlated 
with the FWHM of the lines, resulting in correlation coefficients of 96\%, 
63\%, and 89\% respectively; also a very high degree of correlation. 

\subsection{High resolution spectra}
\begin{figure}
  \begin{center}
	\includegraphics[width=8.5cm]{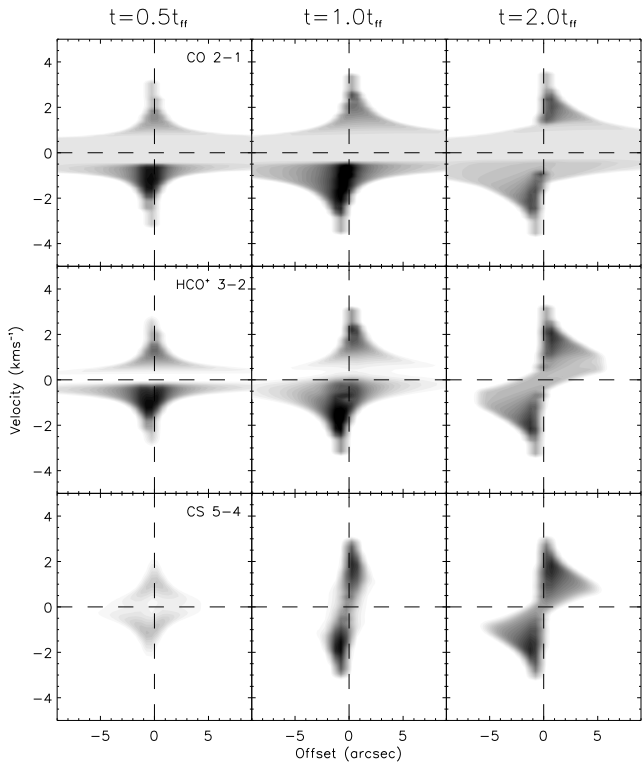}
  \end{center}
  \caption{High resolution position-velocity diagrams of CO 2--1, HCO$^+$ 3--2,
  		   and CS 5--4 at three different stages of the collapse simulation. 
		   These spectra have been convolved with a 1$''$ FWHM Gaussian beam so 
		   that the resolution in these diagrams is comparable to the typical 
		   sub-millimeter interferometer resolution.}\label{pvs}
\end{figure}

With the use of present day and future sub-millimeter interferometers, it is 
possible to obtain spatially resolved observations of protostellar cores and 
protoplanetary disks. Using this kind of instruments, resolutions of 1$''$ or
better are possible. When a resolved spectral image cube is available, it is 
possible to plot the velocity off-set as a function of projected radius by 
contouring the measured intensity in position-velocity space. Figure~\ref{pvs} 
shows PV-diagrams of three time snapshots of each of the three transitions: CO 
2--1, HCO$^+$ 3--2, and CS 5--4. As in Fig.~\ref{specs} we see a clear 
evolution in these diagrams, both in time and with increasing excitation 
energy. In the earliest snapshot, there is emission in almost equal amounts in 
all four quadrants, whereas in the latest snapshot, and most pronounced in the 
CS transition, the emission contours are butterfly-shaped and confined to the 
second and fourth quadrants. The former is typical for radial motion only while 
the latter is associated with rotation.

Although it is to some extent possible to characterize the velocity field that 
results in a particular PV-diagram, it is clear from the t=1.0t$_{ff}$ panels 
in Fig.~\ref{pvs} that the answer one gets depends on the molecule in 
consideration. At t=1.0t$_{ff}$ one would conclude from the CO transition that 
the cloud is contracting whereas from CS the conclusion would be more or less 
pure rotation. The reason for this is that these two transitions are excited at 
different densities and therefore, the PV-diagram shows the dominating velocity 
structure in different regions. This is of course no different from the 
single-dish spectra (Fig.~\ref{specs}), except that the effect is much clearer 
in PV-space.

\section{Results}\label{results}
\subsection{Single-dish simulations}
\begin{figure*}
  \begin{center}
	\includegraphics[width=8.5cm]{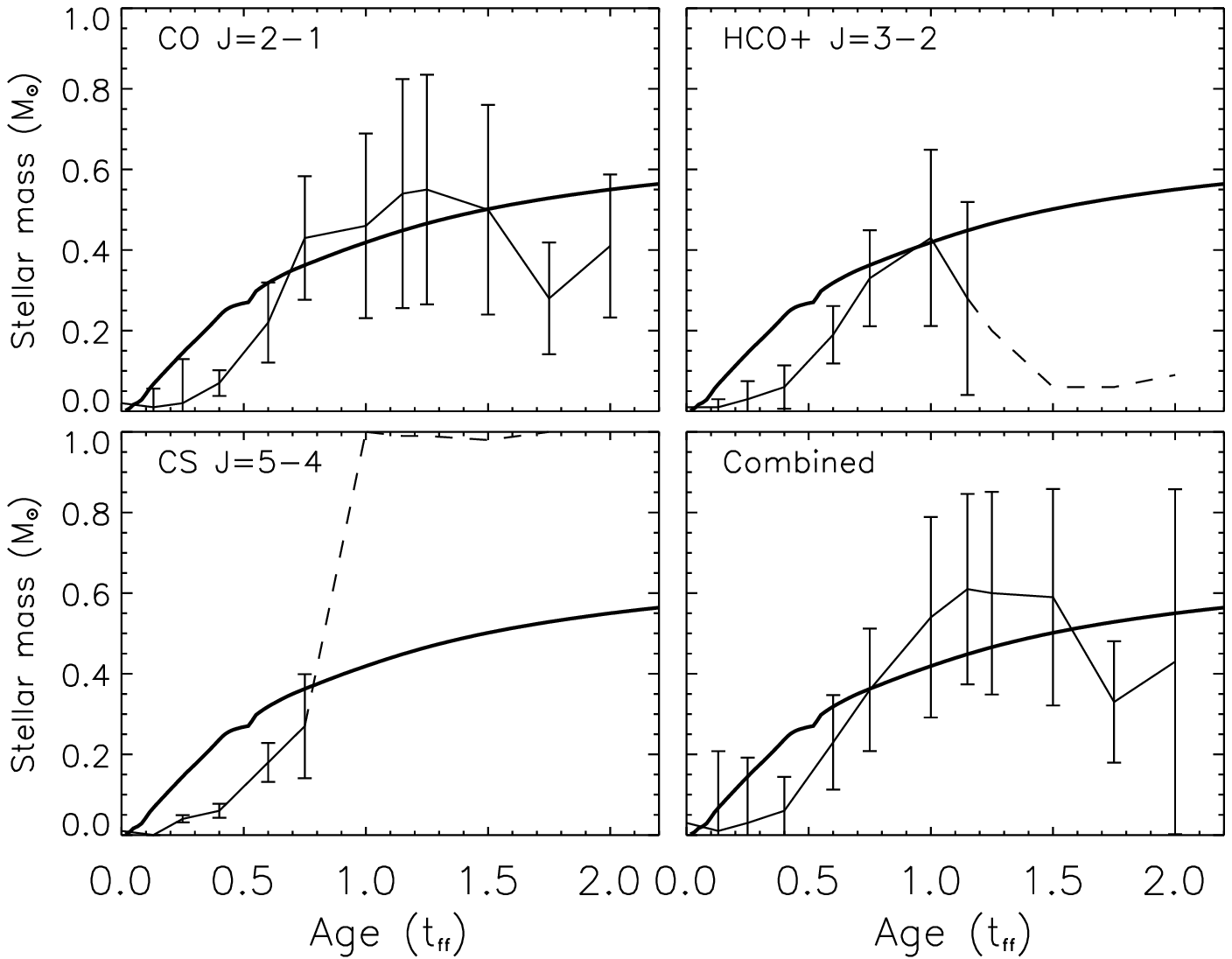}
	\includegraphics[width=8.5cm]{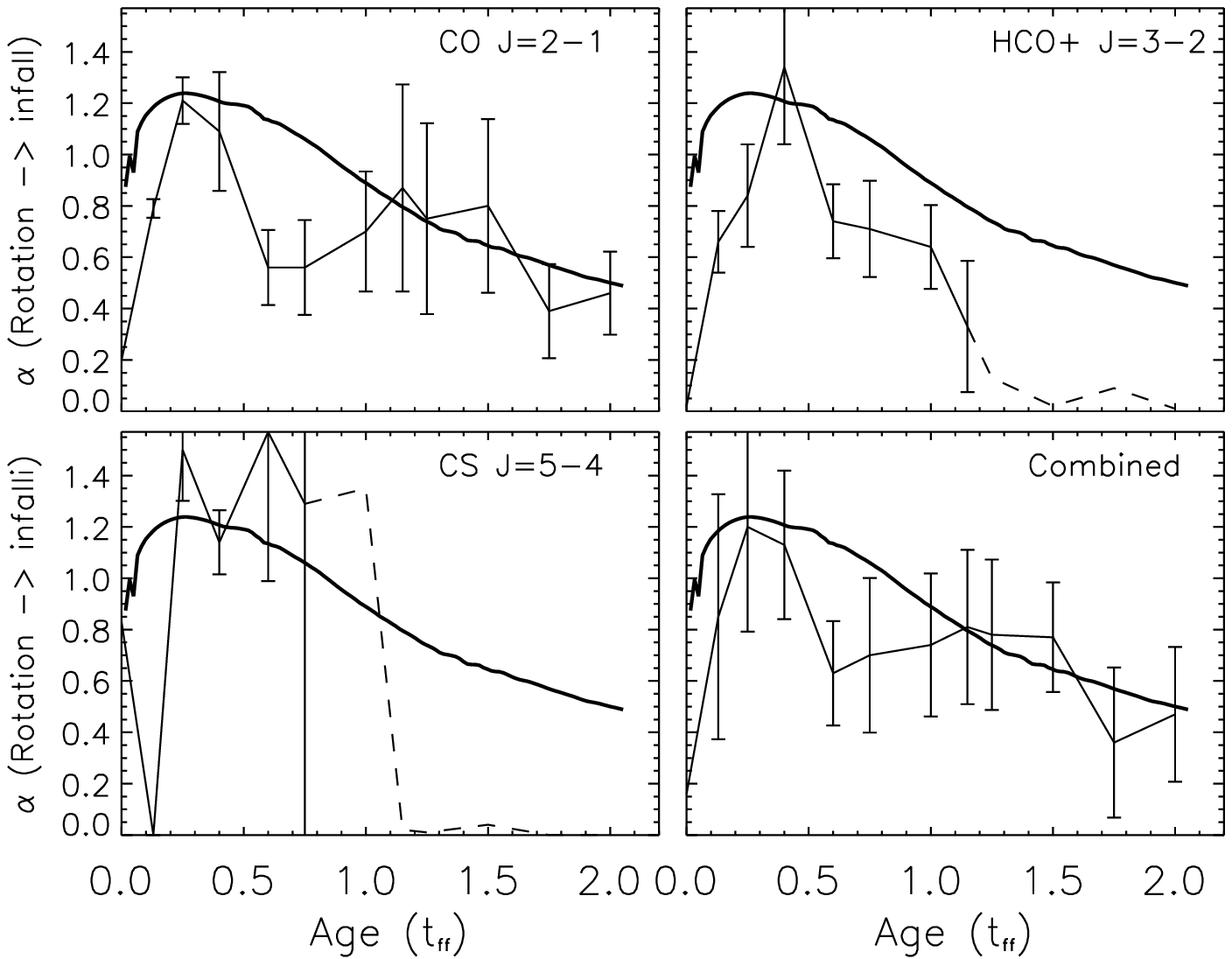}
  \end{center}
  \caption{The result of low-resolution model fits to spectra based on the
           hydrodynamical simulation. The four left panels show the best fit
		   value for the central mass based on CO, HCO$^+$, and CS spectra as
		   well as a fit using all three lines simultaneously. The thick solid
		   line is the values from the simulation and the thin lines connect 
		   the best fit values. The error bars are the width of the $\chi^2$
		   distributions as discussed in sect.~\ref{pikaia}. Dashed lines
		   denotes unconverged models, i.e., cases where the line profile 
		   cannot be reproduced by our model. The four panels to the right are 
		   similar, but for the velocity angle $\alpha$ in radians as defined 
		   by Eq.~\ref{alpha}}.\label{mass-angle}
\end{figure*}

In Fig.~\ref{mass-angle} we present the results of our model fit to the 
spectra from the hydrodynamical simulation using a single spectrum obtained
towards the center of the source in a resolution of 15$''$. This figure shows
the best fit mass (left) and $\alpha$ (right) parameters for the three chosen 
transitions separately, and for a joint fit to all three lines simultaneously. 
First of all we notice that after about 1.0 free-fall time, HCO$^+$ and CS fail
to converge, that is, the model line profile carries no resemblance to the 
line calculated from the hydrodynamical snapshot. However, in the case of the 
mass parameter, for the snapshots earlier the 1.0 free-fall time as well as for 
the entire sequence of CO spectra, we actually get rather accurate parameter 
values, although we do consistently underestimate the mass for 
t$<$1.0 t$_{ff}$. The error bars or robustness of the fit becomes significantly
worse towards the end of the sequence for the mass determination using CO, but
the actual best fit value is close to the correct value. The situation is 
more complicated if we consider the $\alpha$ parameter. Here the CS does not 
show any consistency at any point throughout the sequence. CO and HCO$^+$ are 
considerably better, but the best overall fit is provided by the case where all 
three lines are fit at the same time. 

Generally we see that fitting all three lines at the same time does not give a 
much better fit than the best single transition provides, in this case the CO
line. The fact that CO seems to be the best choice in order to accurately 
describe the velocity field using single-dish observations is somewhat  
contradictory to the result of the correlation calculations in 
sect.~\ref{subsec:lowres}, where the CO profile was seen to be the least 
correlated with the velocity field. According to those numbers, HCO$^+$ and CS
should be able to give much more accurate results the CO. 

The problem is that when using tracers that probe much deeper into the envelope
we start to probe into the regions where our simplified spherical model no 
longer gives an accurate description. Because we use an average density profile 
in order to reproduce the total cloud mass, we automatically get lower column 
densities in the disk region, and this means that lines can get optically thick 
in the hydrodynamical simulation but not in our parameterized model. This is 
the reason why the CS and HCO$^+$ lines are no longer reproducible at t$>$1.0 
t$_{ff}$. Furthermore, in the early snapshots, CS is optically thin and the 
line has thus little structure, which makes it difficult to accurately fix 
$\alpha$.

Interestingly, it seems that a low density tracer such as CO give the most
accurate description of the velocity field even though the CO line never 
actually shows any rotational characteristics (e.g., compare the CO profile at 
t=2.0t$_{ff}$ in Fig.~\ref{specs} to the lower left panel of Fig.~\ref{ideal}),
mainly because CO gets optically thick at radii larger than the outer disk 
radius. Another advantage of using CO as tracer is that it is less affected by
the depletion that occurs in the high density (disk) regions because it gets
optically thick further out. High density tracers such as HCO$^+$ and 
especially CS probe into the region where freeze-out is significant which 
makes the whole process considerably more complicated. On the other hand, 
because CO is excited at low densities and temperatures, the spectrum has a 
much higher chance of being polluted by emission from ambient material. This 
can be hard to identify and if present must be accounted for in the modeling. 
Also, although CO is the best candidate of the three tracers shown here, the 
correlation between the derived velocity field and the real velocity field is 
actually not very high, with most of the points lying between 0.6 and 0.8 
($\approx 45^\circ$). 

\subsection{Interferometric simulations}
Thus we turn to the high-resolution situation, where we use the PV-diagram to
constrain the velocity field parameters. Because the PV-diagram is a 
two-dimensional intensity distribution, comparing model to data is not as 
straight-forward as in the case of using a single spectrum, because there are 
several ways to evaluate a fit in PV-space.

The obvious method is to evaluate the fit by comparing the model PV-diagram to 
the data, pixel by pixel. While this works reasonably well when using 
synthetic, noiseless data which are perfectly aligned with the coordinate axes, 
it may prove very difficult and unreliable using real data, where the position 
angle may not be perfectly well determined. Also, if the model is not 
describing the source geometry very well, especially on small scales, a pixel
by pixel comparison can easily result in a poor fit, although the velocity 
model is actually quite accurate. Nevertheless, we have tried using this 
method, just to see how well it works, when noise is not present.

Another way to obtain a fit is to determine the position of the two maxima of 
the intensity distribution (the blue peak and the red peak) and then use the 
straight line between these two points as a measure. This line is characterized
by three numbers: its length, slope, and orientation in PV-space. This method 
is better when using real data because it works even if the data is quite 
noisy. Unfortunately, the minimum in $\chi^2$-space is less well-defined (the 
fitness landscape is more shallow) using this method, because of the 
degeneracies occurring when using just a linear gradient to evaluate the fit 
and therefore, in most cases, the error bars are quite large.

A third method that we have tried is to use the asymmetry in the PV-diagram as
a constraint. We do this by summing up the emission in each of the four 
quadrants and then evaluate the fit by comparing the ratio of emission in 
quadrant 1 and 4 to the emission in quadrant 2 and 3 as well as the ratio of 
emission in 1 and 2 to 3 and 4. In other words, we measure the left/right and 
top/bottom asymmetry in the PV-diagram. This method has the disadvantage that 
it results in a diagonal ridge in $\chi^2$-space, and thus a further constraint 
is needed. To solve this degeneracy, we use a single-dish line. However, using 
the entire single-dish line profile together with the PV-diagram results again 
in a shallow fitness landscape which we are not interested in. On the other 
hand, we only need to constrain one parameter with the single-dish line to 
resolve the degeneracy, and therefore we mask the center part of the line and 
only fit the line wings. We mask out all channels which lie within the FWHM of 
the line and use the remainder to calculate a $\chi^2$ value which is then 
added to the $\chi^2$ value we get from the PV-diagram.

\begin{figure}
  \begin{center}
	\includegraphics[width=8.5cm]{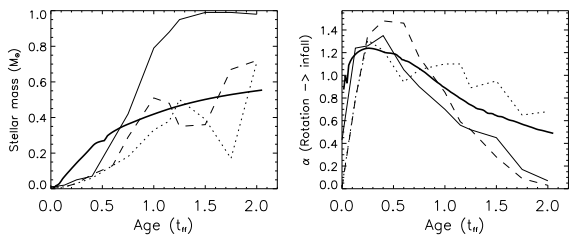}
  \end{center}
  \caption{Similar to Fig.~\ref{mass-angle}, but for HCO$^+$ 3--2 in a 
           resolution of 1$''$, using the PV-diagram to evaluate the $\chi^2$.
		   The three thin line styles correspond to three different method to
		   evaluate the $\chi^2$. The full line corresponds to fitting the 
		   PV-diagram pixel by pixel, the dashed line is the fit obtained from
		   a linear gradient in PV-space, and the dotted line is the 
		   combination of using the PV-diagram and a single-dish line.}
		   \label{hires}
\end{figure}

The result of these three methods are shown in Fig.~\ref{hires}. The three thin
lines correspond to the three methods described above, with the pixel by pixel
method plotted as the full line, the linear gradient as the dashed line and the
emission asymmetry together with the single-dish line wings as the dotted
line. The three methods more or less agree on the results, except for the mass
parameter using the pixel by pixel comparison, which diverges after about 0.5
free-fall times. The main difference between the three methods is the size of 
the error bars (not shown in Fig.~\ref{hires}), which vary quite dramatically,
not only from one method to the other, but also from snapshot to snapshot with
the same method. Generally, the third method described above gives the smallest
errors, except for a few of the earlier snapshots where the linear gradient has
a better defined minimum in $\chi^2$-space. 

\begin{figure}
  \begin{center}
	\includegraphics[width=8.5cm]{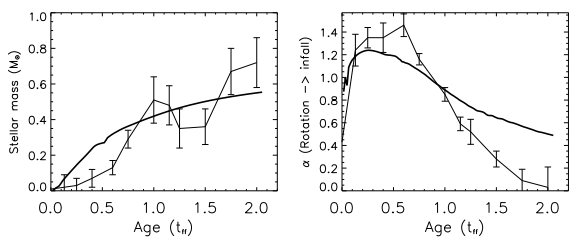}
  \end{center}
  \caption{High-resolution model fit, similar to that of Fig.~\ref{hires}.
   		   Each point shown here together with its associated error bar
		   has been picked among the three different solutions shown in
		   Fig.~\ref{hires} using the smallest error bar as selection 
		   criterion.}\label{hires2}
\end{figure}

Because each of our methods comes with an estimate of how well defined the fit 
is, for each snapshot we can select the method that is most discriminating. 
This is
shown in Fig.~\ref{hires2}, together with the corresponding error bars. The 
resulting curve describes the parameters of the hydrodynamical simulation far 
better than the case of the single-dish result and also better than any of the 
three methods individually shown in Fig.~\ref{hires}. The error bars are also 
very small in almost all of the snapshots which means that the fits are very 
robust or in other words, we can always find a method that gives us acceptably 
small error bars. The main deviation in Fig.~\ref{hires2} is the 
$\alpha$-parameter the later part of the collapse. Here, our model 
overestimates the amount of rotation, near the end by a factor 10. However, the 
graph has a very smooth behavior compared to the best result of the single-dish 
lines, where the solution fluctuates around the correct values. Also, the error 
bars are very small, compared to those of the single-dish result, and this 
means that it should be possible to determine relative ages in a sample of 
objects using the same approach as the one we used to obtain the result in 
Fig.~\ref{hires2}, even though the actual absolute value derived for $\alpha$ 
is incorrect.

\section{Discussion}\label{discussion}
The above analysis has been made using synthetic data which in many respects 
are unrealistic in the sense that they are noiseless and uncontaminated by 
fore- or background material. Furthermore we have disregarded the chemistry, 
using a constant abundance distribution instead. In the following we will 
discuss the effect on our result that using real data will have.

As mentioned in Sect~\ref{model}, freeze-out chemistry is important in 
YSOs~\citep{aikawa2001a,bacmann2002} and this is likely what causes the most 
concern with respect to our results. Freeze-out occurs in regions of high 
density and low temperature, a condition which is present in the outer parts of 
the disk and inner parts of the envelope. As the molecules freeze out onto 
grain surfaces, they no longer contribute to the radiation field of the 
rotational transitions, and if the gas in the entire disk is frozen out we do 
not see any rotational motion at all. There is no doubt that YSOs in general 
are affected by some freeze-out, but the question is to what extent this 
freeze-out affects the spectra.

In a subsequent paper (Brinch et al, in prep) a realistic freeze-out model has 
been added to the hydrodynamical simulation presented in this paper, in order 
to get time-resolved abundance profiles. In that paper only the CO molecule is
considered, but the result may be generalized to include at least also HCO$^+$ 
because the chemistry of this molecule follows that of CO closely. The main 
result of that paper is that the optically thick spectra are not much affected 
throughout the simulation. Freeze-out occurs mainly beyond the $\tau$=3 
isosurface of $^{12}$CO. This is interesting since the $\tau$=3 surface is 
located at the same physical radius no matter whether there is freeze-out 
within this radius or not. This means that in our models too, we never actually 
probe the disk regions directly, and thus we conclude that when we seem to 
measure rotation, we are actually rather measuring lack of infall.

On the other hand, optically thin lines are quite strongly affected by the 
freeze-out zones, because these lines probe the full column density. In 
Sect.~\ref{model} we described how we use an optically thin isotopologue to 
determine the average abundance, and therefore a change in an optically thin 
line due to freeze-out will cause a wrong abundance determination which will 
then propagate to the optically thick lines that are used to obtain the 
velocity field. If freeze-out is present in the data, we can either still try 
to fit it with a constant abundance profile or we can add a simple 
parameterized freeze-out model, following the approach 
of~\citet{jorgensen2005}. In the former situation we derive an average 
abundance which is about a factor of two lower than the correct value for the 
undepleted zones, whereas if we allow for freeze-out in our model as well we 
derive a gas-phase abundance which is about 20\% lower than the correct 
abundance. This should be compared with the situation described in 
Sect.~\ref{model} where we had no freeze-out and we were able to derive the
abundance to within 1\%. However, underestimating the abundance by a factor of 
2 does not make a big difference for the continued analysis and therefore it 
should be safe to use a flat abundance distribution in the model, even though 
the object in question is known to have depletion zones. It is of course also 
always possible to take a non-constant abundance profile into consideration 
when modeling sources for which the chemistry is already 
known~\citep[e.g., as in the case of NGC1333-IRAS2,][]{jorgensen2005a}.

Another source of confusion is contamination from foreground 
material~\citep{boogert2002a}, ambient cloud material or 
outflows~\citep{bally1983}. Because larger scale mapping can identify the 
former of the two, only the outflows are a matter of real concern. Outflow
emission shows up as a broadening and enhancement of the line wings and, if 
present, is typically seen in tracers such as CO and HCO$^+$. Only single-dish 
lines are really affected by these effects, because interferometers typically 
resolve out large scale structure, such as ambient cloud material, and outflow 
regions are easily identifiable and can be masked when making the PV-diagram. 
Foreground material is less of a concern if HCO$^+$ is used, because this 
molecule is only excited at densities higher than what is typical for diffuse 
interstellar foreground layers. For single-dish lines, ambient material may 
shown up as excess emission near the line center which makes it impossible to 
use the line profile to trace the velocity field. If, however, only the line 
wings are used, ambient or foreground material should not be a problem. Thus we 
are left with the outflow contamination of the single-dish lines, which mainly 
affects the line wings.

If a source is known to drive a strong outflow, or if inconsistent results are 
obtained when fitting the model with and without the use of a single-dish line,
it is necessary to either use a different single-dish line that is less 
affected by the outflow or simple to avoid using single-dish lines for that 
particular source. This may result in a less strict result, but there is no way 
to reliably remove the contribution of an outflow from a spectral line unless 
the central mass and to some extent also the velocity field (and thus the shape 
of the line wing) are already known. 

Finally, real data are noisy and this may cause confusion, when fitting the 
single-dish lines. As was mentioned above, using a linear gradient or comparing
the total emission in the four quadrants of the PV-diagram is little affected
by the presence of noise. However, for the single-dish data, it is important to
use strong lines (CO and HCO$^+$ are almost always strong) and integrate long
enough so that the line wings have good single-to-noise. If this is not 
possible, it is again simply a matter of using the PV-diagram without the 
constrain of a single-dish line, at the price of a less certain result.

\section{Conclusion}\label{conclusion}
We have explored the possibility of placing Young Stellar Objects in an 
evolutionary ordering based on their kinematical configurations by fitting a 
simple parameterized model to sub-millimeter observations. We have tested
the feasibility of this by fitting the simple velocity model to a time series 
of synthetic observations with known velocity fields, generated by a 
hydrodynamical simulation. We find that the model reproduces the synthetic 
spectra reasonably well and that the best fit parameters which describe the 
velocity field are in agreement with the velocity field in the hydrodynamical 
simulation. We therefore conclude that it is possible to extract reliable 
information on astronomical objects if we replace the synthetic spectra with 
real observations. 

We find that it is difficult, though not impossible, using single-dish data 
alone, due to the shallowness of the $\chi^2$-space, but feasible if 
interferometric data are used, especially when combined with one or more 
single-dish lines. Using the latter option, we find that the central dynamical 
mass can be determined to within 20\%. However, we have not been able to 
establish a single ``best way'' to obtain the parameters throughout the 
collapse. We find that the best result is obtained by trying several different 
methods to constrain the fits, and then pick the result of the one which shows 
the most peaked $\chi^2$-space. That said, most of our best constrained points 
were obtained using a combination of low-resolution and high-resolution and 
this is the preferred starting point.  

We further conclude that the molecular species used are not crucial for the 
result, as long as the line wings are clearly defined and uncontaminated in the
case of the single-dish lines, and also that the transition is optically thick.
For the high-resolution data, signal-to-noise is the main concern, so strong
lines are preferred (CO, HCO$^+$, etc.). 

Due to the uncertainty of the absolute time scale of the collapse calculated by 
our relatively simple hydrodynamical simulation, we cannot yet 
establish an absolute age of a given object, but relative ages among a sample 
of objects can be obtained, provided that the velocity field of a collapsing 
cloud evolves similar to that of the simulation. Indeed, if an age calibration 
could be made, for example using the chemical properties of one or more 
objects, this could prove a powerful method to place a large number of young 
stars in an evolutionary sequence, because of the relatively straight-forward 
observations needed to perform the analysis. The hydrodynamical simulation 
presented in this paper is not required in order to analyze real objects using 
our method. However, with an improved and more realistic hydrodynamical scheme, 
it might be possible to calibrate the evolution of the velocity field so that 
an absolute time scale of star formation can be established. \\

\noindent \emph{Acknowledgments}  CB is partially supported by the European 
Commission through the FP6 - Marie Curie Early Stage Researcher Training 
programme. The research of MRH is supported through a VIDI grant from the 
Netherlands Organization for Scientific Research.\\

\bibliographystyle{aa}
\bibliography{/home/brinch/papers/references}

\end{document}